\documentclass{article}
\usepackage{spconf,amsmath,graphicx}
\usepackage[table]{xcolor}
\usepackage{xcolor}
\usepackage{booktabs}
\usepackage{multirow}
\usepackage{array}
\usepackage{svg}

\newcolumntype{C}[1]{>{\raggedleft\arraybackslash}p{#1}}

\title{Hierarchical Emotion Prediction and Control in Text-to-Speech Synthesis}
\name{
Sho Inoue\textsuperscript{1,2}, 
Kun Zhou\textsuperscript{3}, 
Shuai Wang\textsuperscript{2,$\dag$}, and 
Haizhou Li\textsuperscript{1,2}\thanks{$\dag$: Corresponding author} \thanks{This research is supported by China NSFC project under Grant No.62271432, Shenzhen Science and Technology Research Fund, Fundamental Research Key Project under Grant No.JCYJ20220818103001002, and Internal Project of Shenzhen Research Institute of Big Data under grant No.T00120220002}
}

\address{
\textsuperscript{1} School of Data Science, The Chinese University of Hong Kong, Shenzhen (CUHK-Shenzhen), China\\
\textsuperscript{2}Shenzhen Research Institute of Big Data, Shenzhen, China\\
\textsuperscript{3}Speech Lab of DAMO Academy, Alibaba Group, Singapore
}
\begin{document}
\ninept
\maketitle
\begin{abstract}
It remains a challenge to effectively control the emotion rendering in text-to-speech (TTS) synthesis. Prior studies have primarily focused on learning a global prosodic representation at the utterance level, which strongly correlates with linguistic prosody. Our goal is to construct a hierarchical emotion distribution (ED) that effectively encapsulates intensity variations of emotions at various levels of granularity, encompassing phonemes, words, and utterances. During TTS training, the hierarchical ED is extracted from the ground-truth audio and guides the predictor to establish a connection between emotional and linguistic prosody. At run-time inference, the TTS model generates emotional speech and, at the same time, provides quantitative control of emotion over the speech constituents. Both objective and subjective evaluations validate the effectiveness of the proposed framework in terms of emotion prediction and control. 

\end{abstract}

\section{Introduction}\label{sec:intro}

Emotional text-to-speech (TTS) aims to synthesize realistic emotional speech from the text \cite{schroder2001emotional}. 
Neural TTS systems have achieved significant improvement in generating natural-sounding voices. However, they usually struggle to express the exact emotions~\cite{triantafyllopoulos2023overview, schuller2018age}. 
Emotional TTS seeks to address this problem for natural human-computer interaction \cite{pittermann2010handling,kun2022emotion}.

TTS models not only the human vocal system but also the prosodic variations presented in human speech~\cite{tan2021survey}. Speech prosody affects both the syntactic and semantic interpretation of an utterance (``linguistic prosody'') and also embodies the speaker's emotional state (``emotional prosody'')~\cite{hirschberg2004pragmatics}. Although these two types of prosody are functionally independent, they share common acoustic characteristics~\cite{belyk2014perception}. Emotional TTS considers a modulation of both linguistic and emotional prosody, which further raises two questions: 1) how to model emotional prosody with lexical content; 2) how to control emotion rendering over linguistic units. In this paper, we aim to address these two challenges.


Previous emotional TTS studies characterize emotions as a global feature of the utterance~\cite{kun-intensity,kun-mix}. TTS models learn to associate emotional styles with explicit labels during training. 
For instance, some studies utilize a reference encoder to encode the emotional styles into a global vector such as global style tokens~\cite{gst}. 
To enhance user control, researchers employ relative attributes~\cite{parikh2011relative} to control the intensity level of the output emotion~\cite{zhu2019controlling,kun-mix}. 
We observe a lack of focus from previous methods on modeling the relationships between emotional prosody and the semantic representation of the utterance, as well as providing quantitative emotion control of different linguistic units within a sentence.

In this paper, we propose a novel approach to predict and control emotion rendering over texts. The proposed framework automatically predicts emotional content from text and allows users to control the emotion rendering over different segments.
Our contributions are highlighted as follows:

\vskip-1.0em

{
\setlength{\leftmargini}{10pt} 
\begin{itemize}
\item We introduce hierarchical emotion distribution (ED) predictor\footnote{\textbf{Implementation}: https://github.com/shinshoji01/Text-Hierarchical-ED}, a quantifiable emotion predictor that deduces the hierarchical emotion distribution at different granularity, solely from the text. 
\item During training, the hierarchical ED predictor is guided to predict hierarchical ED from the semantic representations produced by a BERT \cite{bert}-based linguistic encoder.
Hierarchical ED can be automatically predicted from a given text input and manually modified during inference;
\item Our method demonstrates enhanced emotion expressiveness in emotional speech synthesis and offers efficient and flexible emotion control across various segmental levels.
\end{itemize}
}

The rest of this paper is organized as follows: In Section 2, we introduce related works. Section 3 describes our proposed methodology. In Section 4, we introduce our experiments and analyze the results. Section 5 concludes our study. 

\begin{figure*}[h]
		\centering
		\centerline{\includegraphics[width=14.6cm]{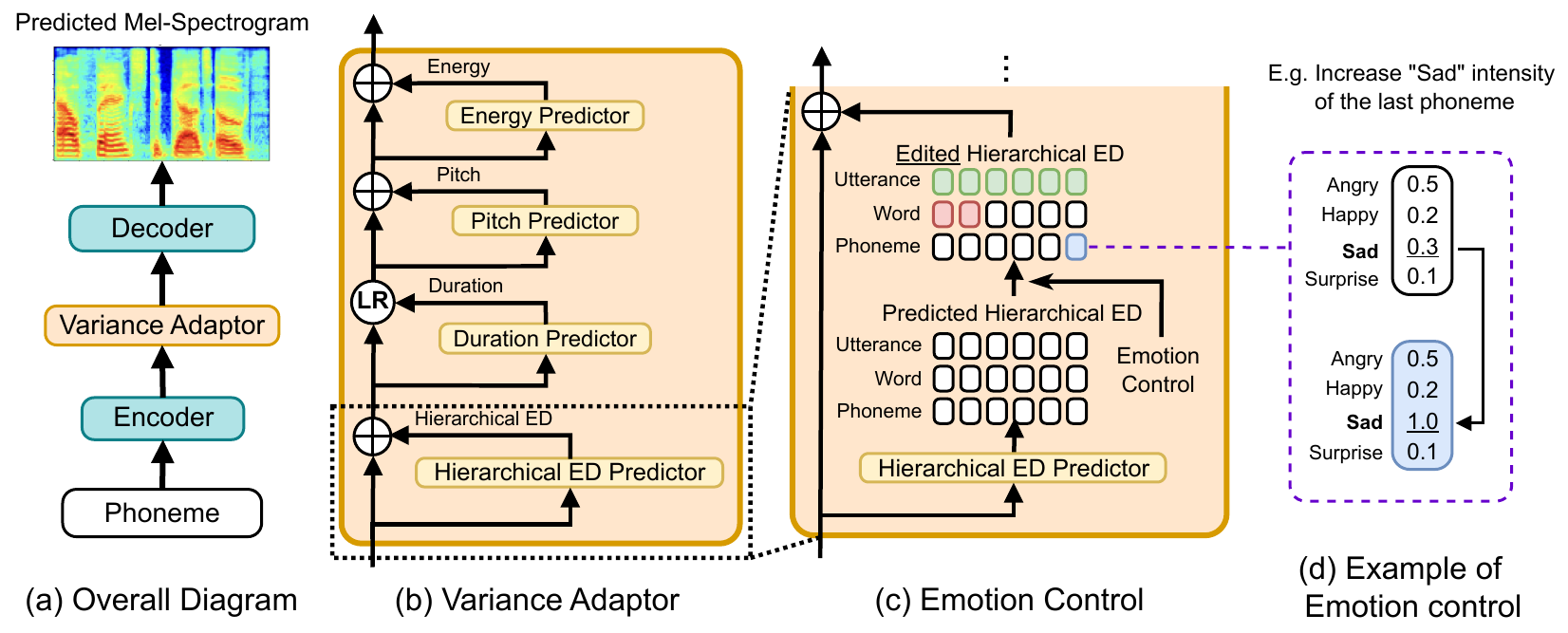}}
		\vspace{-0.3cm}
	\caption{
 System diagrams of (a) Model architecture; (b) Variance adaptor integrating a Hierarchical Emotion Distribution (ED) predictor; (c) Emotion control diagram; (d) An example of emotion control.
}
\vspace{-0.4cm}
	\label{fig:diagram}
\end{figure*}

\section{Text-based Prosody Prediction for TTS}\label{sec:related_works}
A spoken utterance carries both linguistic prosody and emotional prosody. One technical challenge in TTS is to predict the exact prosody for an input text.   
FastSpeech2~\cite{fastspeech2} explicitly utilizes ground-truth pitch, energy, and duration for training, and predicts these features solely from text during inference. \cite{fspbert} leverages BERT \cite{bert} to enhance
FastSpeech2. 
Other studies predict implicit prosody labels exclusively from text~\cite{textGST,plmdn,hieTTS}. \cite{plmdn} employs a GMM-based mixture density network to predict phoneme-level prosodic embeddings, while \cite{bert1,bert2} utilizes BERT \cite{bert} to extract emotion embeddings from the transcript.
Despite much progress, the prosody prediction techniques lack the necessary fine-grain control over the speech constituents. Therefore, it is difficult to control the emotion rendering.

We observe that the prior studies lack focus on emotional prosody modeling over linguistic units. In this paper, we focus on predicting and controlling emotional prosody in different segmental levels (phoneme, word, and utterance). For example, we want the model to know where to put emotional emphasis when producing an utterance (`emotion prediction'), or to manipulate the emotion of an utterance to begin with joy but end with sorrow (`emotion control'). Such a technique is required in conversational speech synthesis. 


        



\section{Proposed Methods}
The overall diagram of our proposed framework is shown in Fig.\ref{fig:diagram}(a), which is based on the FastSpeech2 \cite{fastspeech2} structure. Given a phoneme sequence, our proposed framework predicts the hierarchical ED and prosodic variants (duration, pitch, and energy) and synthesizes the Mel-Spectrogram. 
We first replace the linguistic encoder with a BERT-based encoder~\cite{bert} to enhance its knowledge of semantic information in the sentences.  
We further introduce a novel hierarchical ED predictor that quantifies emotion intensity in a hierarchical manner and allows users to assign and adjust the intensity of emotions over different linguistic units. The proposed framework is expected to: 1) produce a more natural emotional outcome by modeling both emotional and linguistic prosody, and 2) enhance the linguistic-wise emotion control of the emotional TTS system.

\subsection{Hierarchical Emotion Distribution Extractor}\label{sec:SED_extractor}
{
We first present the concept of hierarchical emotion distribution (ED), a novel emotion representation that incorporates continuous emotion intensity labels across different granularity levels including phonemes, words, and utterances. Specifically, we design an extractor to generate hierarchical ED from any datasets containing emotional speech and emotion labels, making it a versatile tool for this purpose. 
}
Our implementation is inspired by the study of relative attributes, initially proposed in computer vision \cite{parikh2011relative} and later extended to speech processing \cite{kun-intensity,kun-mix,msemotts}. We first frame emotion style in speech as an attribute, then construct a ranking function to quantify the relative presence of an emotion.
We obtain the relative attributes from the ranking functions and normalize them to the range $[0,1]$, where a larger value indicates a higher intensity of emotion. 

We initially apply the Montreal Forced Alignment (MFA)~\cite{mfa} for the alignment of words and phonemes. 
We then use OpenSMILE~\cite{opensmile} to extract an 88-dimensional feature set for each segment level of an audio (phoneme, word, and utterance). Then, we use pre-trained ranking functions to obtain emotion intensity labels for speech segments. 

We define the ranking function as 
$f(x_i) = w \cdot x_i + b$,
where $w$ and $b$ denote the weight vector and bias, respectively. The parameters are optimized using the support vector machine's objective function for binary classification \cite{svm}:
\begin{align*}
\min_{w, b} \frac{1}{2} \|w\|^2 + C \sum_{i=1}^{n} \max(0, 1 - y_i (w \cdot x_i + b))
\end{align*}
Here, $x_i$ represents the acoustic features of the $i$-th training sample, $C$ the regularization parameter, $y_i$ is the training label, for example, to train the ranking function for ``Happy'':
$y_i = \begin{cases}
    +1   & (i\in H)\\
    -1  & \text{otherwise}
\end{cases}$
where $H$ denotes the set of ``Happy'' samples.
We utilize different segments (phonemes, words, utterances) of each audio as training samples.
The extractor with pre-trained relative functions could obtain the emotion intensity labels for a speech segment, which serve as the training labels for the text-based emotion prediction training described in Section 3.2.

\subsection{Predicting Hierarchical Emotion Distribution from Text}

Most datasets group utterances into several emotion categories \cite{zhou2021seen}. However, the intra or inter-sentence emotion intensity variations are often overlooked. We note that intra-sentence emotion intensity variations are associated with different segmental levels of an utterance, which calls for the design of a text-based hierarchical emotion prediction model.

We construct a linguistic encoder equipped with a broad understanding of semantic information. 
We replace FastSpeech2's encoder with a BERT-based encoder, renowned for its proficiency in natural language processing~\cite{bert}.
We aim to predict hierarchical emotion distribution solely from text.
We design a hierarchical ED predictor and incorporate it into the variance adaptor, as shown in Fig.\ref{fig:diagram}(b). The ground-truth hierarchical ED is obtained by the extractor from the speech signal at the phoneme, word, and utterance level, as described in Section 3.1. During TTS training, the variance adaptor learns the hierarchical emotion distribution, duration, pitch, and energy from the linguistic embedding sequentially. In this way, the TTS framework associates the hierarchical emotional prosody and prosodic variants over linguistic representations.  

\subsection{Quantifiable Emotion  Control}
Our proposed framework facilitates quantifiable emotion control during inference, as shown in Fig.\ref{fig:diagram}(c). The BERT-based encoder first encodes a phoneme sequence into a linguistic embedding.
The variance adaptor then predicts hierarchical ED at three different granularity levels (phoneme, word, and utterance) from the linguistic embedding. The emotion distribution is illustrated in Fig.\ref{fig:diagram}(d), where each value represents a predicted intensity of an emotion type. The users can change or assign those values to create a desired emotion rendering at three granularity levels.

\subsection{Comparison with Related Work}

In this work, we introduce a TTS framework with quantifiable emotion prediction and control. Our proposal is similar to~\cite{msemotts} but differs in many ways. First, 
\cite{msemotts} only considers utterance-level emotion and its phoneme-level intensities. During inference, \cite{msemotts} keeps a consistent emotion type within the entire utterance, only allowing for alterations in intensity. Our model considers the hierarchical nature of emotions, and model emotion distribution at three granularity levels (phoneme, word, utterance), allowing for manipulating the intensity of each emotion in any granularity level.

\section{Experiments and Results}
\subsection{Model Architecture}
We utilize FastSpeech2~\cite{fastspeech2} as our backbone framework, comprising a text encoder, variance adaptor, and decoder.
The text encoder is based on a transformer network~\cite{transformer}, which transforms a phoneme sequence into a linguistic embedding. 
The variance adaptor is composed of Hierarchical ED, duration, pitch, and energy predictors.
For hierarchical ED prediction, we reuse the pitch predictor's structure, which quantizes the pitch of each phoneme to 256 values through feed-forward networks. 
A transformer-based decoder then synthesizes a mel-spectrogram from the output of the variance adaptor.
We substitute FastSpeech2's linguistic encoder with a BERT-based\cite{bert} encoder that accepts phoneme inputs. Both the transformer encoder's hidden layers and attention heads are set to 8. 
We follow the same optimizer~\cite{adam} and the learning rate scheduler in FastSpeech2.
The batch sizes for TTS training and BERT pretraining are 128 and 64, respectively.


\subsection{Experimental Setup}
We conduct experiments using three datasets: Blizzard Challenge 2013 dataset~\cite{blizzard}, Emotion Speech Dataset (ESD)~\cite{esd}, and BookCorpus~\cite{bookcorpus}. 
We train our TTS model on the Blizzard dataset, which is derived from audiobooks 
and contains expressive speech data with abundant prosody variance without any emotion labels. 
ESD~\cite{esd} contains emotional speech data grouped into 5 categories: Neutral, Sad, Angry, Happy, and Surprise.
We use English recordings from 10 speakers. We randomly select 20 samples per speaker and emotion to train the ranking functions for the Hierarchical ED extractor. 
The BookCorpus dataset~\cite{bookcorpus} is sourced from over 11k books from a variety of genres, which is utilized to pre-train the linguistic encoder. 
We extracted about 1/10 of the entire dataset, equivalent to approximately 7M sentences. 


We pre-train the BERT-based linguistic encoder over 3 epochs in Masked Language Modeling tasks. Once integrated into the TTS framework, we fine-tune it for 100k iterations, while keeping the encoder's parameters fixed. Then, we update the entire architecture for an additional 400k iterations. The losses for TTS tasks include (1) L1 loss from mel-spectrograms and 
(2) MSE loss from prosodic predictions. The vocoder is HiFiGAN~\cite{hifigan} pre-trained on the Blizzard dataset.

As a fair comparison, we implemented the following systems: 
\setlength{\leftmargini}{5pt} 
\begin{itemize}
    \item \textbf{FastSpeech}: We implemented FastSpeech2 \cite{fastspeech2}, where the variance adaptor predicts duration, pitch, and energy from the text;
    \item \textbf{Proposed Method}: Our proposed framework described in Section 3, leverages a BERT-based linguistic encoder and is equipped with a hierarchical emotion distribution predictor;
    \item \textbf{Proposed Method w/o ED Predictor}: Our proposed framework that only leverages a BERT-based linguistic encoder;
    
    \item \textbf{Proposed Method w/o BERT}: Our proposed framework is only equipped with a hierarchical ED predictor.
\end{itemize}

\noindent
Note that the linguistic encoder in  FastSpeech comprises 11.9M trainable parameters, contrasting with the BERT-based encoder's larger size of 34.2M parameters.
Next, we present the experimental results on the effectiveness of our proposed framework in terms of speech quality, speech expressiveness, and emotion controllability, by comparing with the baselines.

\begin{table}[t]
\vskip -0.15in
\caption{
Speech quality evaluation results of 1) mean opinion score (MOS) and 2) Mel-cepstral distortion (MCD).
}
\vspace{-2mm}
\label{table:mos}
\begin{center}
\scalebox{0.85}{
\begin{tabular}{l||c|c}
\toprule
Frameworks& MOS $\uparrow$ & MCD $\downarrow$ \\
\toprule
Ground Truth & 4.183$\pm$0.268 & ---\\
FastSpeech2 & 3.946$\pm$0.282 & 4.945$\pm$0.155 \\ \midrule
Proposed & 3.933$\pm$0.213 & 4.866$\pm$0.143\\
\hspace{1em} - ED Predictor & 3.895$\pm$0.229 & 4.892$\pm$0.148 \\
\hspace{1em} - BERT & 3.607$\pm$0.272 & 4.974$\pm$0.137 \\
\bottomrule
\end{tabular}
}
\end{center}
\vskip -0.3in
\end{table}

\begin{table}[t]

\caption{
Best-worst scaling (BWS) test result, where the value represents the ratio of preferences from the evaluators ($\%$). Red and green colors represent the selection ratio of the least similar audio and the most similar audio, respectively.
}
\vspace{-0.4cm}
\label{table:bws}
\begin{center}
\scalebox{0.74}{
\begin{tabular}{C{0.7cm}C{0.7cm}|C{1.5cm}C{1.5cm}|C{1.1cm}C{1.1cm}|C{0.7cm}C{0.7cm}}
\toprule
\multicolumn{2}{c|}{FastSpeech2} & \multicolumn{2}{c|}{Proposed w/o ED Predictor} & \multicolumn{2}{c|}{Proposed w/o BERT} & \multicolumn{2}{c}{Proposed}\\
\midrule
Worst & Best &Worst& Best &Worst & Best &Worst &Best\\
\midrule
\cellcolor{red!86}{43$\%$} & \cellcolor{green!24}{12$\%$} & \cellcolor{red!36}{18$\%$} & \cellcolor{green!52}{26$\%$} & \cellcolor{red!46}{23$\%$} & \cellcolor{green!54}{27$\%$} & \cellcolor{red!32}{16$\%$} & \cellcolor{green!70}{35$\%$} \\
\bottomrule
\end{tabular}
}
\end{center}
\vskip -0.35in
\end{table}

\begin{table*}[!t]
\caption{The average prosody change ratio in response to increasing intensity from low (0.0) to high (1.0)
evaluated on Blizzard and ESD datasets. 
The cell hue denotes the heat-mapped values, consistent across the dataset and the prosody.
(``Word and Phoneme'' is a combination of word-level and phoneme-level control.)}
\label{table:control_score}
\begin{center}
\scalebox{0.8}{
\begin{tabular}{cc||c|c|||c|c|||c|c|||c|c|||c|c|||}
\toprule
& & \multicolumn{2}{c|||}{Duration} & \multicolumn{2}{c|||}{Pitch (mean)} & \multicolumn{2}{c|||}{Pitch (std)} & \multicolumn{2}{c|||}{Energy (mean)} & \multicolumn{2}{c|||}{Energy (std)}\\
& & Blizzard & ESD & Blizzard & ESD & Blizzard & ESD & Blizzard & ESD & Blizzard & ESD \\
\midrule
\multirow{4}{4em}{Utterance} & Angry &\cellcolor{green!1}{0.012} & \cellcolor{red!1}{-0.018} & \cellcolor{green!5}{0.015} & \cellcolor{green!8}{0.019} & \cellcolor{red!3}{-0.036} & \cellcolor{red!3}{-0.049} & \cellcolor{green!3}{0.017} & \cellcolor{green!1}{0.007} & \cellcolor{green!14}{0.055} & \cellcolor{green!6}{0.032}\\
& Happy &\cellcolor{red!1}{-0.010} & \cellcolor{red!0}{-0.013} & \cellcolor{green!21}{0.060} & \cellcolor{green!19}{0.042} & \cellcolor{red!14}{-0.146} & \cellcolor{red!6}{-0.085} & \cellcolor{green!12}{0.055} & \cellcolor{red!6}{-0.033} & \cellcolor{green!8}{0.033} & \cellcolor{green!3}{0.017}\\
& Sad &\cellcolor{red!0}{-0.001} & \cellcolor{green!0}{0.004} & \cellcolor{green!10}{0.028} & \cellcolor{green!7}{0.017} & \cellcolor{red!15}{-0.156} & \cellcolor{red!2}{-0.036} & \cellcolor{green!3}{0.017} & \cellcolor{green!4}{0.022} & \cellcolor{red!4}{-0.017} & \cellcolor{red!1}{-0.006}\\
& Surprise &\cellcolor{green!0}{0.004} & \cellcolor{green!0}{0.005} & \cellcolor{red!10}{-0.029} & \cellcolor{green!10}{0.023} & \cellcolor{green!19}{0.195} & \cellcolor{green!4}{0.056} & \cellcolor{red!14}{-0.065} & \cellcolor{green!3}{0.017} & \cellcolor{green!1}{0.006} & \cellcolor{red!0}{-0.005}\\
\hline\hline
\multirow{4}{4em}{Word and Phoneme} & Angry &\cellcolor{green!69}{0.449} & \cellcolor{green!35}{0.447} & \cellcolor{green!12}{0.035} & \cellcolor{green!25}{0.054} & \cellcolor{green!29}{0.297} & \cellcolor{green!50}{0.674} & \cellcolor{red!6}{-0.030} & \cellcolor{green!0}{0.001} & \cellcolor{green!99}{0.373} & \cellcolor{green!100}{0.524}\\
& Happy &\cellcolor{green!15}{0.098} & \cellcolor{green!12}{0.160} & \cellcolor{green!100}{0.276} & \cellcolor{green!100}{0.213} & \cellcolor{red!0}{-0.001} & \cellcolor{green!6}{0.093} & \cellcolor{red!35}{-0.159} & \cellcolor{red!74}{-0.385} & \cellcolor{red!4}{-0.015} & \cellcolor{red!5}{-0.031}\\
& Sad &\cellcolor{green!100}{0.648} & \cellcolor{green!100}{1.259} & \cellcolor{green!3}{0.010} & \cellcolor{green!5}{0.012} & \cellcolor{green!3}{0.038} & \cellcolor{green!99}{1.346} & \cellcolor{red!100}{-0.443} & \cellcolor{red!100}{-0.518} & \cellcolor{red!28}{-0.106} & \cellcolor{green!66}{0.347}\\
& Surprise &\cellcolor{red!1}{-0.013} & \cellcolor{red!1}{-0.014} & \cellcolor{green!30}{0.083} & \cellcolor{green!90}{0.192} & \cellcolor{green!100}{1.023} & \cellcolor{green!56}{0.765} & \cellcolor{red!39}{-0.174} & \cellcolor{green!55}{0.287} & \cellcolor{red!17}{-0.067} & \cellcolor{green!26}{0.137}\\
\hline\hline
\multirow{4}{4em}{Word} & Angry &\cellcolor{green!6}{0.039} & \cellcolor{green!7}{0.089} & \cellcolor{green!30}{0.084} & \cellcolor{green!50}{0.108} & \cellcolor{red!10}{-0.108} & \cellcolor{green!4}{0.064} & \cellcolor{green!32}{0.143} & \cellcolor{green!45}{0.234} & \cellcolor{green!35}{0.133} & \cellcolor{green!35}{0.184}\\
& Happy &\cellcolor{green!15}{0.103} & \cellcolor{green!13}{0.173} & \cellcolor{green!38}{0.107} & \cellcolor{green!41}{0.089} & \cellcolor{green!17}{0.174} & \cellcolor{green!15}{0.210} & \cellcolor{red!9}{-0.042} & \cellcolor{red!17}{-0.088} & \cellcolor{green!14}{0.056} & \cellcolor{green!14}{0.076}\\
& Sad &\cellcolor{green!22}{0.145} & \cellcolor{green!22}{0.278} & \cellcolor{green!27}{0.076} & \cellcolor{green!42}{0.090} & \cellcolor{red!9}{-0.093} & \cellcolor{green!9}{0.134} & \cellcolor{red!18}{-0.084} & \cellcolor{red!19}{-0.100} & \cellcolor{green!5}{0.021} & \cellcolor{green!31}{0.164}\\
& Surprise &\cellcolor{green!6}{0.043} & \cellcolor{green!4}{0.060} & \cellcolor{green!24}{0.069} & \cellcolor{green!50}{0.107} & \cellcolor{green!29}{0.301} & \cellcolor{green!24}{0.336} & \cellcolor{green!8}{0.039} & \cellcolor{green!48}{0.249} & \cellcolor{green!13}{0.051} & \cellcolor{green!19}{0.104}\\
\hline\hline
\multirow{4}{4em}{Phoneme} & Angry &\cellcolor{green!19}{0.127} & \cellcolor{green!14}{0.177} & \cellcolor{red!23}{-0.064} & \cellcolor{red!23}{-0.050} & \cellcolor{green!11}{0.117} & \cellcolor{green!19}{0.262} & \cellcolor{red!21}{-0.096} & \cellcolor{red!26}{-0.136} & \cellcolor{green!82}{0.309} & \cellcolor{green!60}{0.316}\\
& Happy &\cellcolor{red!5}{-0.034} & \cellcolor{green!0}{0.009} & \cellcolor{green!17}{0.050} & \cellcolor{green!12}{0.027} & \cellcolor{green!91}{0.941} & \cellcolor{green!19}{0.267} & \cellcolor{red!17}{-0.077} & \cellcolor{red!39}{-0.207} & \cellcolor{green!5}{0.022} & \cellcolor{red!12}{-0.064}\\
& Sad &\cellcolor{green!33}{0.216} & \cellcolor{green!41}{0.519} & \cellcolor{red!41}{-0.115} & \cellcolor{red!33}{-0.071} & \cellcolor{red!44}{-0.458} & \cellcolor{green!2}{0.036} & \cellcolor{red!62}{-0.279} & \cellcolor{red!74}{-0.384} & \cellcolor{green!7}{0.028} & \cellcolor{green!40}{0.210}\\
& Surprise &\cellcolor{red!8}{-0.052} & \cellcolor{red!4}{-0.060} & \cellcolor{red!3}{-0.011} & \cellcolor{red!40}{-0.086} & \cellcolor{green!66}{0.678} & \cellcolor{green!44}{0.600} & \cellcolor{red!38}{-0.170} & \cellcolor{green!9}{0.048} & \cellcolor{red!2}{-0.010} & \cellcolor{green!16}{0.085}\\
\bottomrule
\end{tabular}
}
\end{center}
\vskip -0.3in
\end{table*}

\subsection{Results and Discussion}\label{sec:results}
We conduct both objective and subjective evaluations in terms of speech quality, emotion expressiveness, and emotion controllability. As for subjective evaluation, 15 subjects participated in our listening experiments, where each of them listened to a total number of 80 samples guided by detailed instructions. We present speech samples on our demo page\footnote{\textbf{Speech Demo}: https://shinshoji01.github.io/Text-Sequential-ED-Demo/}.


\subsubsection{Speech Quality}
We first evaluate our proposed framework with the baseline models in terms of the overall speech quality through mean opinion score (MOS). MOS is a subjective test to evaluate the audio quality, where listeners are asked to rate the audio on a scale from 1 to 5 with a 0.5 increment. A higher MOS score represents better speech quality. As illustrated in Table~\ref{table:mos}, our model's quality is comparable to the baseline models without ED predictor. The integration of both ED predictor and BERT contributes to the overall speech quality.

We also calculate Mel-cepstral distortion (MCD) to objectively evaluate speech distortion between synthesized and ground-truth speech signals, where a smaller value of MCD represents a smaller distortion and indicates a better speech quality. As shown in Table~\ref{table:mos}, our proposed method outperforms all the other models, which shows a promising performance of overall speech quality. 

\subsubsection{Speech Expressiveness}\label{sec:mcd}

We further conduct experiments to evaluate the speech expressiveness of the synthesized audio. 
We conduct the best-worst scaling (BWS) test, where the subjects listen to four speech samples alongside the reference audio and choose the speech sample that has the most and the least similarity with the reference in terms of speech expressiveness. 
As presented in Table~\ref{table:bws}, most of the evaluators choose our proposed framework (35\%) as the `best', and only 16\% choose it as the worst, which consistently outperforms all the other models. The BWS results show the effectiveness of our proposed framework to predict speech expressiveness from the text. 

%
%

\subsubsection{{Emotion Controllability}}

We examine prosody alteration by studying how the model responds when users manually control the emotion intensity. 
We define controllability as the consistency between the prosody change in synthesized audio and the expected prosodic behaviors derived from prior speech analysis studies~\cite{esd,emoprosody}.
We employ a metric that quantifies the ratio of prosodic changes between the lowest and the highest emotion intensities. This metric is calculated using the formula $(p_h - p_l) / p_l$, where $p_h$ and $p_l$ represent the prosodic feature values for the highest (1.0) and the lowest (0.0) intensities, respectively. 
Essentially, this score reflects the alteration in emotion prosody of the synthesized audio in terms of the increase in emotion intensity.

For each utterance, we randomly select either 5 words or 20 phonemes to adjust their intensities. The duration data is sourced from the predictions made by FastSpeech2, while pitch and energy values are derived from OpenSMILE~\cite{opensmile}. Additionally, we conduct experiments involving fine-tuning a text-to-speech (TTS) model with ESD dataset~\cite{esd} to assess its capability of controlling emotions.

Table~\ref{table:control_score} shows the average prosody change ratio for each emotion and segment. A higher value suggests a more pronounced prosody change. 
We observed that we achieve the highest controllability at word and phoneme-level, but notably lower controllability at the utterance level. Moreover, we discovered that combining word and phoneme-level controls yields an average effect similar to conducting word-level and phoneme-level emotion controls independently.
On Blizzard and ESD datasets, we note certain consistent correlations between emotions and acoustic features. Specifically, we observe a positive correlation between ``Angry'' and ``Duration/Energy (std)'', ``Happy'' and ``Pitch (mean)'', ``Sad'' and ``Duration'', and ``Surprise'' and ``Pitch''. In contrast, there is a negative correlation between ``Sad'' and ``Energy (mean)''. 
To clarify, a positive correlation between ``Happy'' and ``Pitch (mean)'' represents an increased intensity of ``Happy'' emotion indicating an elevated speech pitch.
In summary, the results in Table~\ref{table:control_score} are consistent with prior studies in speech emotion analysis~\cite{esd,emoprosody}, 
underscoring the efficacy of our model in emotion control.
It is worth mentioning that the ESD dataset exhibits more alignments with the literature than the Blizzard dataset, such as the presence of a positive correlation between ``Surprise'' and ``Energy (mean)'', and ``Sad'' and ``Energy (std)''.
These differences may stem from the emotional prosody that is explicitly exhibited in ESD.




\section{Conclusion}

We introduce an emotional TTS framework with hierarchical emotion prediction and control. Our proposed framework leverages a pre-trained BERT-based encoder to produce meaningful linguistic representations from the text inputs. We design a novel emotion distribution extractor that learns hierarchical emotion information from the speech signal. We train a variance adaptor to predict hierarchical emotion distribution, duration, pitch, and energy from the linguistic representation in a sequential manner. Our model allows users to control the emotion rendering at different granularity levels at run-time.
Both subjective and objective evaluations demonstrate our model's proficiency in emotion prediction and control. 

\bibliographystyle{IEEEbib}
\bibliography{strings,refs}
\end{document}